\documentclass{aa}
\usepackage{graphicx}
\usepackage{natbib}
\bibpunct{(}{)}{;}{a}{}{,}
\begin{document}
\title{Altair's inclination from line profile analysis\thanks{Based on
    observations carried out at Observatoire de Haute-Provence (OHP),
    at the German-Spanish Astronomical Centre, Calar Alto, and
    retrieved from the Elodie Archive at Observatoire de
    Haute-Provence (OHP)}}

  \author{A. Reiners\inst{1,2}
    \and
    F. Royer\inst{3,4}}
  
  \offprints{A. Reiners}
  
  \institute{
   Astronomy Department, University of California, Berkeley, CA 94720 \\ \email{areiners@astron.berkeley.edu}
   \and
   Hamburger Sternwarte, Universit\"at Hamburg, Gojenbergsweg 112, 21029 Hamburg, Germany
   \and
   Observatoire de Gen\`eve, 51 chemin des Maillettes, 1290 Sauverny, Switzerland \\ \email{frederic.royer@obs.unige.ch}
   \and
   GEPI, CNRS UMR 8111, Observatoire de Paris, 5 place Janssen, 92195 Meudon cedex, France
  }
  
  \date{\today}
  
  \abstract{We present a detailed spectroscopic study of line
    broadening in the A7IV-V star Altair. In a wavelength region
    covering 690\AA\ we reconstruct the overall broadening profile
    taking into account more than 650 spectral lines.  From the
    broadening profile we determine the projected rotational velocity
    $v\,\sin{i}$, derive an upper limit for the equatorial velocity
    $v$ from the shape of the profile and search for signatures of
    differential rotation. Our redetermined value of $v\,\sin{i}$ is
    ($227 \pm 11$)\,km\,s$^{-1}$. Measuring the first two zeros of the
    Fourier transformed broadening profile yield no signatures of
    differential rotation. We derive that Altair is seen under an
    inclination angle higher than $i = 68^{\circ}$ and it rotates at
    $v < 245$\,km\,s$^{-1}$ or slower than 53\% of breakup velocity on
    a $1\sigma$ level.   }
  
\maketitle

\section{Introduction}
As the 12th brightest star in the sky, Altair ($\alpha$~Aql, HR~7557,
HD~187\,642) has been extensively studied in the literature. Recently
\citet{Bui_04} list most of its physical parameters in their
introduction. Altair is an A7IV-V main sequence star known to be a
fast rotator with a projected rotational velocity $v\sin i =
217$~km\,s$^{-1}$ \citep{Royer02}.  At a distance of $5.14$~pc
\citep{Hip}, it is one of the nearest bright stars, allowing
\citet{vanBelle} to measure the oblateness of the star using
interferometric observations.  They derived an axial ratio $a/b =
1.140 \pm 0.029$ and a value for the $v\sin i$ of $210 \pm
13$~km\,s$^{-1}$.  Interferometry also allows to make constraints on
stellar inclination -- a parameter usually not available from standard
spectroscopic techniques. Their determination of the inclination $i$
of the rotational axis ranges from $30$\degr\ to $90$\degr\ within
1$\sigma$ \citep[ Fig.~5]{vanBelle}.

Spectroscopic techniques are usually not sensitive for inclination in
``classical'' case of spherical stars. On the contrary, for fast
rotators, the stellar surface is distorted by centrifugal forces and
line profiles depend on the angle between the rotational axis and the
line of sight. Moreover, in such objects effective temperature varies
with gravity over the stellar surface: this is the gravity darkening
\citep{VZl24}. 

Scrutinizing the shape of individual lines, \citet{Gulliver94}
determined the inclination angle of Vega from spectroscopic data. In a
more general context, \citet{Reiners03a} shows the effects of
inclination and gravity darkening on line profiles of fast rotators in
the Fourier domain. He showed that in spectra of fast rotators the
unprojected rotational velocity $v$ is measurable directly from the
line shape. It is the aim of this paper to put stronger constraints in
deriving the inclination of Altair from \textit{spectroscopic}
measurements by measuring its (unprojected) equatorial velocity and
the $v\sin i$.

\section{Data}

\subsection{Altair data}

Altair has been observed on October 1st 2003 with the \'ELODIE
\'echelle spectrograph \citep{Bae_96} at Observatoire de
Haute-Provence. The wavelength coverage spans from 3850 to 6800 \AA,
with a resolving power of $42\,000$. Five single exposures have been
taken, details about the observational data are listed in
Table~\ref{ObsData}.

The orders of the \'echelle spectrum have been merged using the method
by \citet{ErrNoh02}.  The barycentric radial velocity correction
varies of about $0.025$~km\,s$^{-1}$ in the short time interval of the
observations, and meanwhile, the peculiar radial velocity of Altair is
not expected to change significantly.  The comparison of wavelength
positions for telluric lines (O$_2$) insures that no significant
instrumental wavelength shift occurred between the consecutive
exposures, and the merged spectra are therefore simply co-added.

\begin{table*}

  \caption{\label{tab:Stars}The collected spectra are listed for each stellar target, 
    giving the corresponding spectrograph, the resolving power, the 
    Julian date (HJD) and signal-to-noise ratio (S/N) for each observation.
    The S/N for \'ELODIE spectra is estimated at $5550$~\AA. Values of 
    $v\,\sin{i}$ are from \cite{Royer02} given in km\,s$^{-1}$.}
  
  \label{ObsData}
  \begin{tabular}{llcrcrr}
    \hline
    \hline
    \noalign{\smallskip}
    Object     & Sp.Type & Instrument & $\lambda/\Delta \lambda$ & HJD & S/N & $v\,\sin{i}$\\
    \noalign{\smallskip}
    \hline
    \noalign{\smallskip}
    Altair     & A7IV/V &   \'ELODIE   &   $42\,000$   &  2452914.2574  &  180 & $\sim 220$\\
    & &              &               &  2452914.2600  &  200 &\\
    & &              &               &  2452914.2628  &  280 &\\
    & &              &               &  2452914.2655  &  230 &\\
    & &              &               &  2452914.2683  &  250 &\\
    \noalign{\smallskip}
    \hline
    \noalign{\smallskip}
    HD 27\,819  & A7IV/V &   \'ELODIE   &   $42\,000$  & 2450499.4221 & 180 & $\sim 40$\\
    \noalign{\smallskip}
    \hline 
    \noalign{\smallskip}
    $\gamma$ Boo  & A7III/IV &   FOCES      &  $40\,000$  & 2452336.0736 & 300 & $\sim 130$\\
    \noalign{\smallskip}
    \hline
    \noalign{\smallskip}
    HD 118\,623  &  A7III &  FOCES      &  $40\,000$  & 2452332.1416 & 300 & $\sim 210$\\
    \noalign{\smallskip}
    \hline
  \end{tabular}
\end{table*}

\subsection{Calibration data}

We used spectra of three stars of similar spectral type to calibrate
our measurements in Altair's spectrum. Spectra observed with the FOCES
spectrograph at the 2.2m telescope at the German-Spanish Astronomical
Centre, Calar Alto, were taken in February 2002, for details on the
reduction cf. \cite{Reiners03}. The spectrum of HD~27\,819 was taken
from the \'ELODIE
Archive\footnote{\texttt{http://atlas.obs-hp.fr/elodie/}}. The
calibration stars are listed in Table\,\ref{tab:Stars}.

\section{Continuum normalization}

Analyzing spectral line shapes severely depends on the determination
of the continuum flux. In spectra of stars with rotation rates as high
as Altair's, spectral lines are severely blended and the flux hardly
reaches the continuum in any spectral region. Thus, normalizing
Altair's spectrum by eye-guided continuum fitting would always assume
too low a continuum which significantly affects the shape of the line
profiles.

Thus we took high quality spectra of other stars of similar spectral
types, some of them having considerably lower rotation rates; in their
spectra the continuum is nicely visible. From the FOCES campaign we
took two stars; HD~118\,623 is an A7III with a rotational velocity
very similar to Altair. $\gamma$~Boo has a rotational velocity of
$v\,\sin{i} \sim 130$\,km\,s$^{-1}$, i.e.  significantly lower than
that of Altair and HD~118\,623. In the spectrum of $\gamma$~Boo flux
reaches continuum level in some regions and continuum fitting is
possible.

\begin{figure}
 \centering
    \resizebox{\hsize}{!}{\includegraphics[angle=-90]{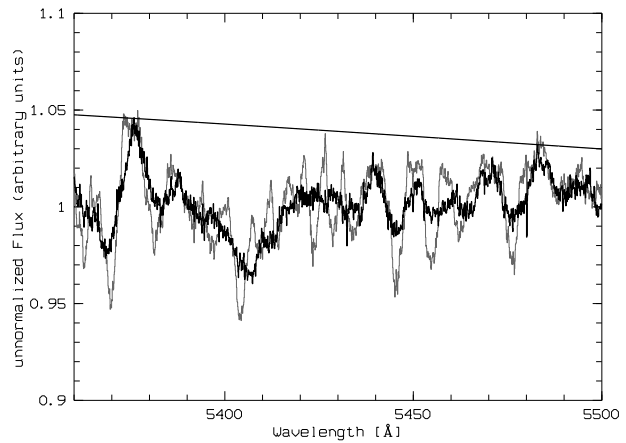}}
    \caption{\label{plot:unnormalized} Small region of the unnormalized 
      spectra of $\gamma$~Boo (grey) and HD~118\,623 (black). The fit
      to the continuum assumed is overplotted and is in good agreement
      with both spectra.}
\end{figure}

The spectrum of $\gamma$~Boo was normalized by fitting the regions
where flux reaches continuum level. We also used this spline fit to
normalize the spectrum of HD~118\,623 which was taken during the same
observing campaign in the same instrument setup. The coincidence of
their unnormalized spectra is shown in Fig.\,\ref{plot:unnormalized},
giving reason for using identical normalization functions.

\begin{figure}
 \centering
    \resizebox{\hsize}{!}{\includegraphics[angle=-90]{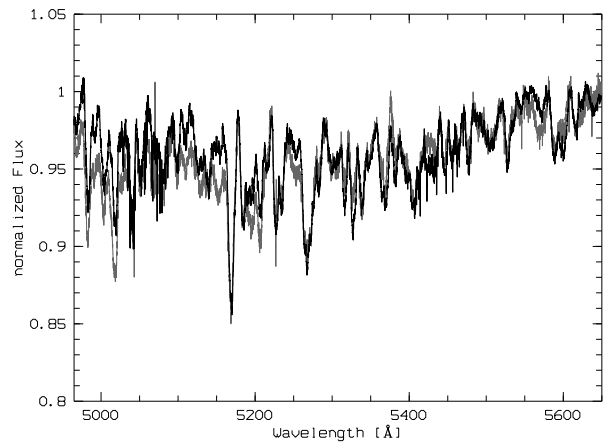}}
    \caption{\label{plot:Altair} Normalized spectra of HD~118\,623 (grey) and
      Altair (black). The whole regions used to derive Altair's
      broadening function is shown.}
\end{figure}

With a normalized spectrum of HD~118\,623, a star being similar to
Altair with respect to spectral type \textbf{and} projected rotational
velocity, we could then normalize the spectrum of Altair by matching
its overall spectral shape to that of HD~118\,623.  Both normalized
spectra are shown in Fig.\ref{plot:Altair} in the whole wavelength
region we used in the following.

\section{Reconstructing the global broadening function}

Spectra of stars rotating as fast as Altair are completely dominated
by rotational broadening.  Temperature broadening and turbulent
motions are about two orders of magnitude smaller than rotation.
Rotation affects all spectral lines in the same characteristic manner.
However, the assumption we have to make is that the line profiles of
individual lines do not significantly (at least not systematically)
vary across the stellar disk.  For individual lines this is not true,
but using hundreds of lines of different species, variations among
different lines are expected to level out. What one gets is a profile
which reflects the \textbf{mean} rotational broadening affecting
\textbf{all} lines. This method is in contrast to studying individual
lines \citep[e.g.,][]{Gulliver94} and line-specific variations with
temperature or gravity are not taken into account. For the case of
Altair, systematic effects are not expected to be as severe as, e.g.,
for the case of Vega. We will discuss that point in
Sect.\,\ref{Temperature}.  However, individual line profiles are
severely blended in Altair's spectrum making studies of individual
lines difficult and we think that recovering line broadening from as
many lines as possible yields the most reliable information about
stellar rotation.

The reconstruction of Altair's mean broadening profile is done by a
deconvolution algorithm. The algorithm searches the function that
gives the best fit to the data when being convolved with a given
template. Thus the template is required to be a precise spectrum of
the star without rotation (neglecting other broadeners). Since no such
template is available we usually try to simultaneously fit the line
strengths, i.e., equivalent widths, of all spectral lines
\citep[cp.][]{Reiners03}. In the case of Altair, however, blending of
the large amount of spectral lines due to high rotational velocity is
so severe, that simultaneously fitting the broadening function and
line strengths is not possible. We thus did not determine the template
from the spectrum of Altair but from a spectrum of a star of similar
spectral type.

\begin{figure}
 \centering
    \resizebox{\hsize}{!}{\includegraphics[angle=-90]{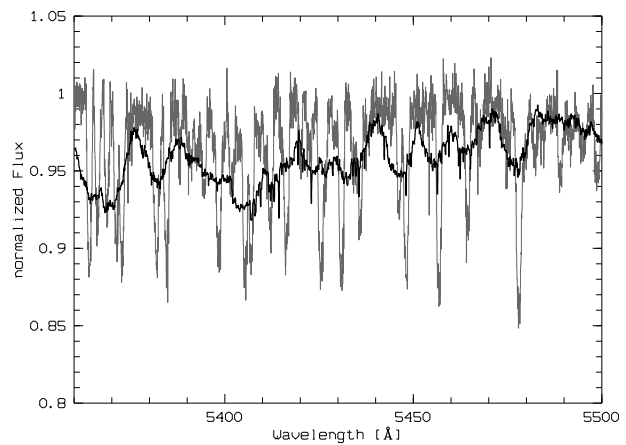}}
    \caption{\label{plot:HD27819} Comparison of the normalized spectra of HD~27\,819
      (grey line) and Altair (black line) in a small spectral window.}
\end{figure}

In spectra of slower rotators with less severe blending,
simultaneously fitting the broadening function and line strengths is
achievable. From the \'ELODIE-Archive we took a spectrum of
HD~27\,819, a A7IV-V star rotating at $v\,\sin{i} \sim
40$\,km\,s$^{-1}$.  Normalization was made by eye using the clearly
visible continuum. A comparison of the normalized spectra of Altair
and HD~27\,819 is shown in Fig.\,\ref{plot:HD27819}; spectral
similarity is clearly visible. For this ``slow'' rotator HD~27\,819
the broadening function and line strengths were derived iteratively
from the spectrum using our deconvolution algorithm
\citep[cp.][]{Reiners03}. We used the whole wavelength region between
$\lambda = 4960$\,\AA\, and $\lambda = 5650$\,\AA. In this spectral
region all lines are dominated by rotation, i.e., hydrogen lines are
not contained. Roughly 650 spectral lines were used in the fit.
Wavelength information was taken from the Vienna Atomic Line Database
\citep[VALD]{VALD}.

\begin{figure*}
  \includegraphics[height=\hsize,width=7cm,angle=-90,clip=]{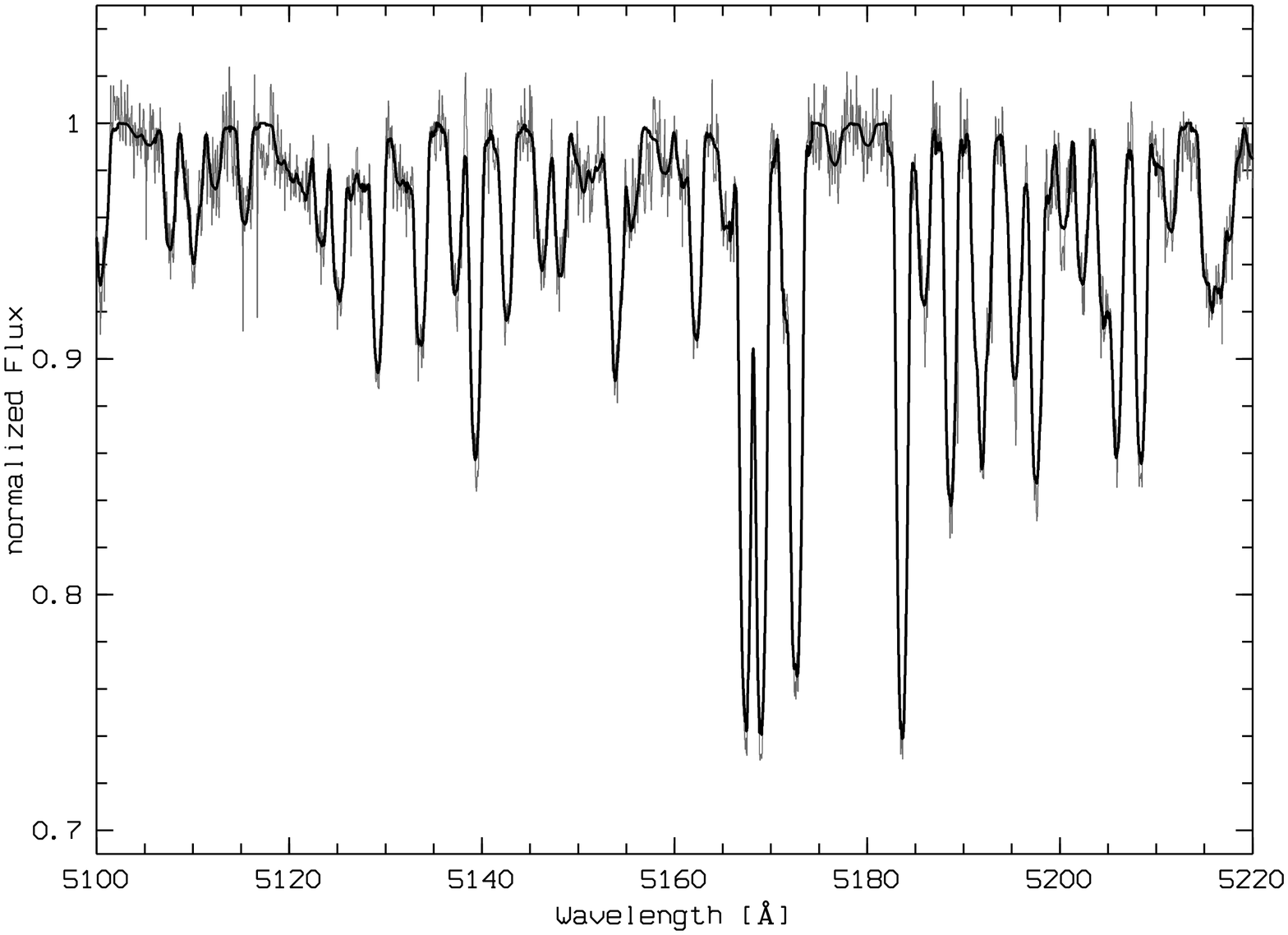}

  \vspace{-3.4cm}

  \hspace{2.8cm}\includegraphics[height=5.2cm,width=2.5cm,angle=-90,clip=]{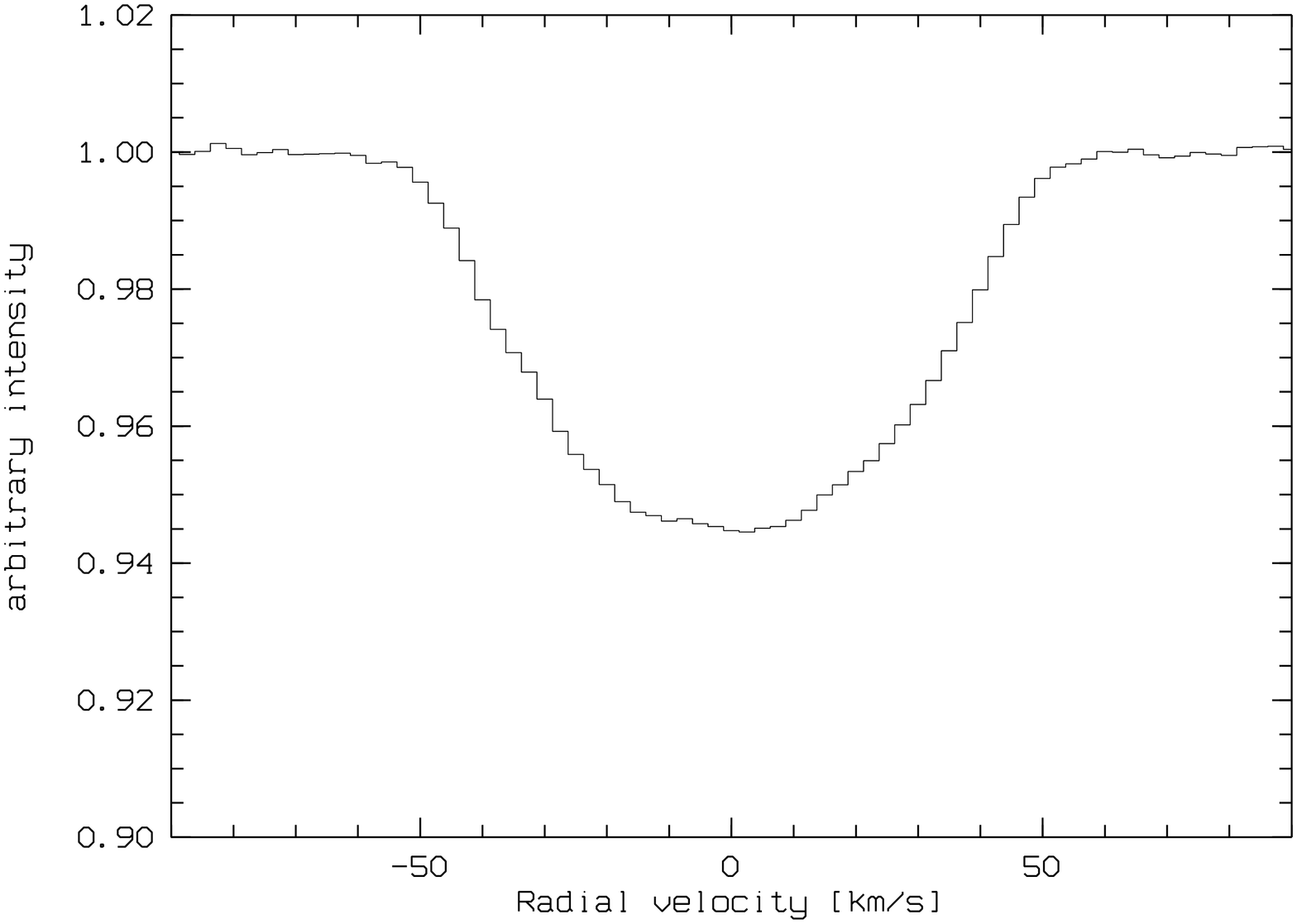}
  \vspace{.8cm}
  \caption{\label{plot:HD27819_fit}Data (grey) and fit
    (black) achieved from our deconvolution algorithm for the case of
    HD~27\,819. Fit quality is typical for the whole wavelength
    region. The deconvolved ``global'' profile is plotted in the
    inset. }
\end{figure*}

The result of this procedure is shown in Fig.\,\ref{plot:HD27819_fit}
where we plot the result of convolving the recovered broadening
function -- shown in the inset -- with the optimized template.  Note
that each pixel of the broadening function is a free parameter in the
algorithm and that the broadening function resembles nicely symmetric
rotational broadening (from the profile we derive a $v\,\sin{i}$ of
46\,km\,s$^{-1}$). The data (grey line) is reproduced in great detail;
since individual lines are not severely blended ambiguities between
the shape of the broadening function and line strengths are very
small.

\begin{figure*}
  \includegraphics[height=\hsize,width=7cm,angle=-90,clip=]{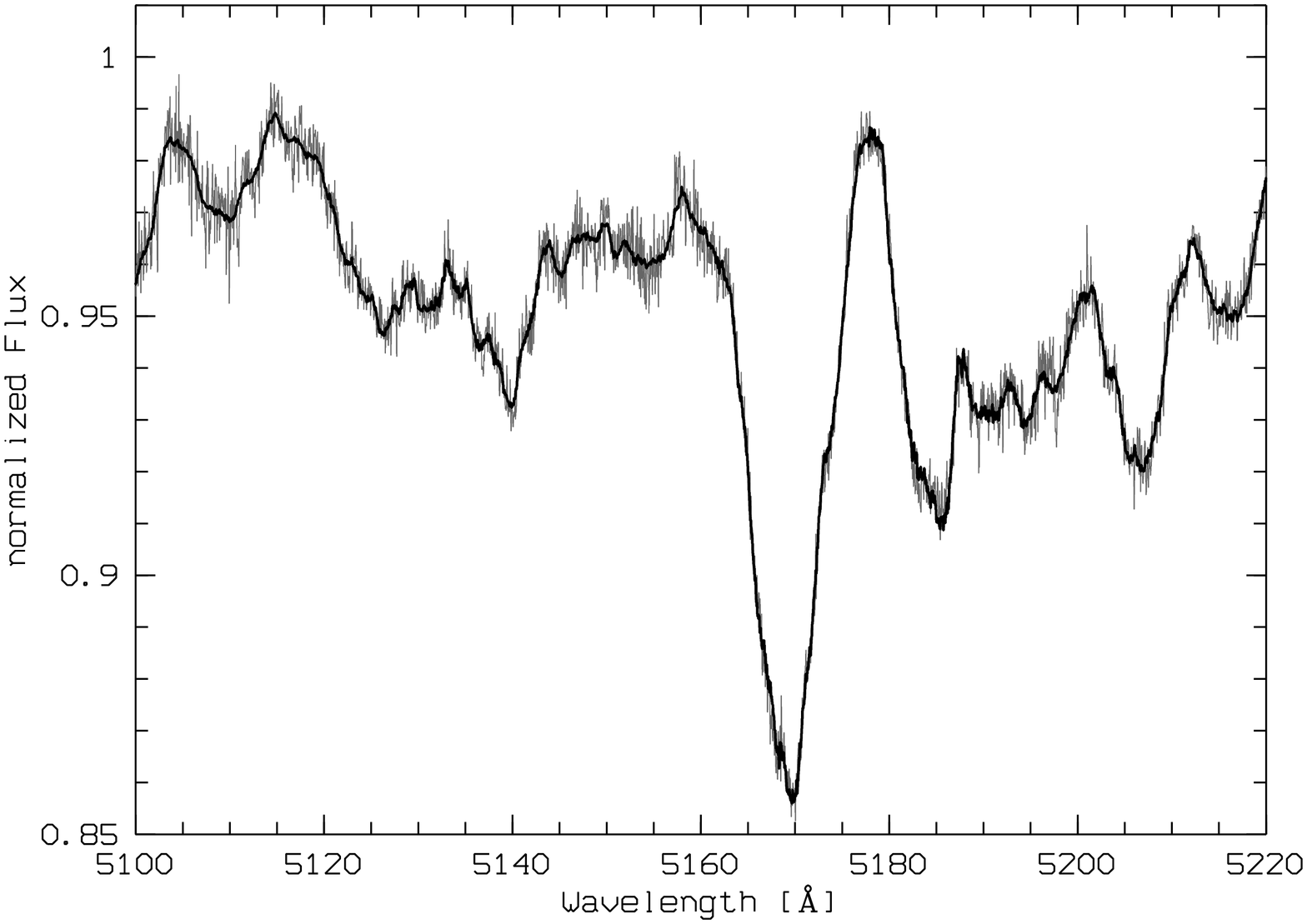}

  \vspace{-3.4cm}

  \hspace{2.8cm}\includegraphics[height=5.2cm,width=2.5cm,angle=-90,clip=]{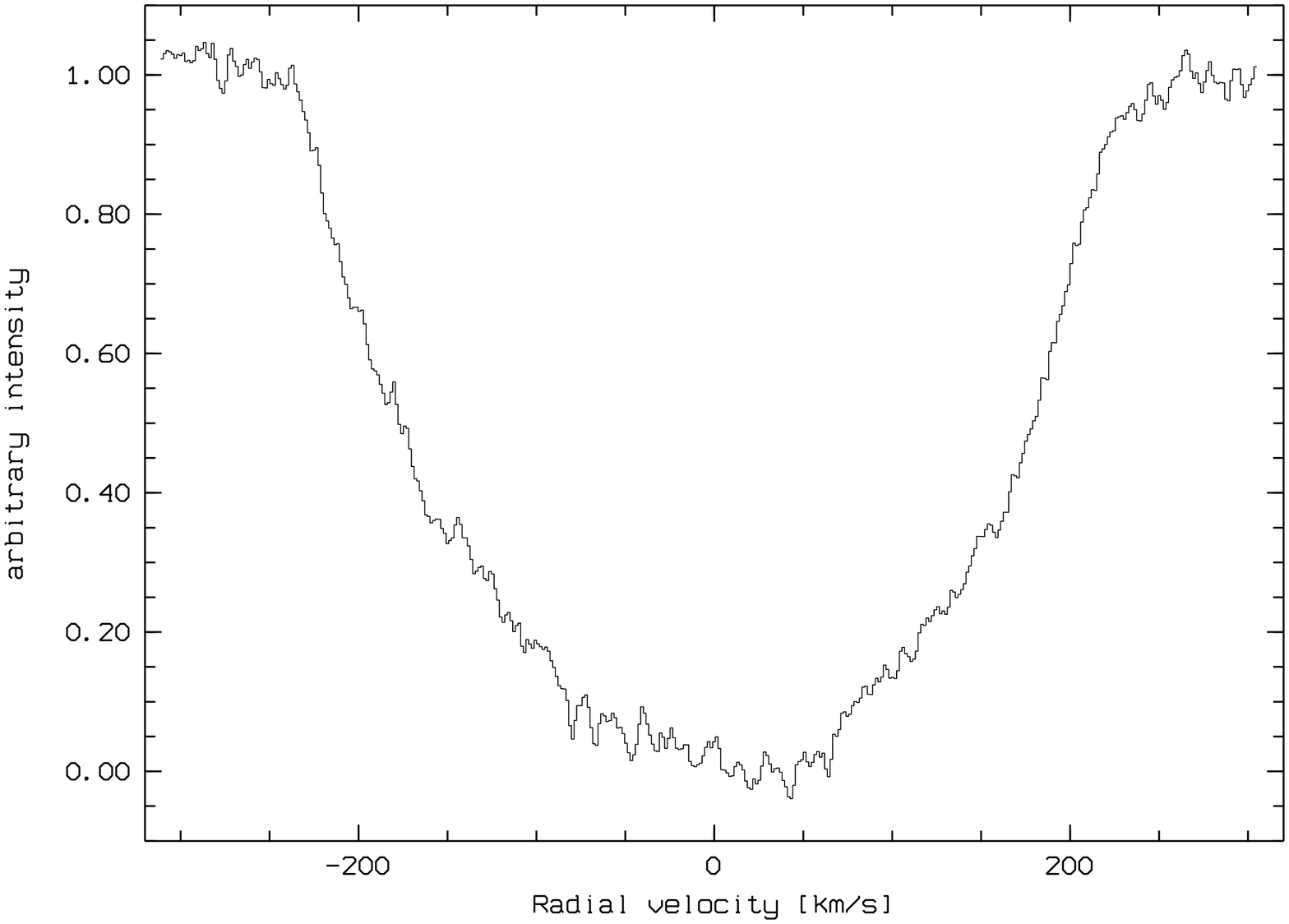}
  \vspace{.8cm}
  \caption{\label{plot:Altair_fit}Data (grey line) and achieved fit
    (black line) of Altair. Fit quality is typical for the whole
    wavelength region. The deconvolved ``overall'' profile is plotted
    in the inset. }
\end{figure*}

For the deconvolution of Altair's broadening profile we now used the
same optimized template, i.e., the same line strengths as derived for
HD~27\,819. Note that the observed spectrum of HD~27\,819 could not be
used as a template since it already is broadened by
$~46$\,km\,s$^{-1}$ and a convolution of two rotational broadening
functions does not yield a third one. From the normalized spectrum of
Altair we finally derived the profile that gives the smallest
chi-square when convolved with this delta-like template.

\begin{figure*}
 \centering \mbox{
    \resizebox{.5\hsize}{!}{\includegraphics[angle=-90]{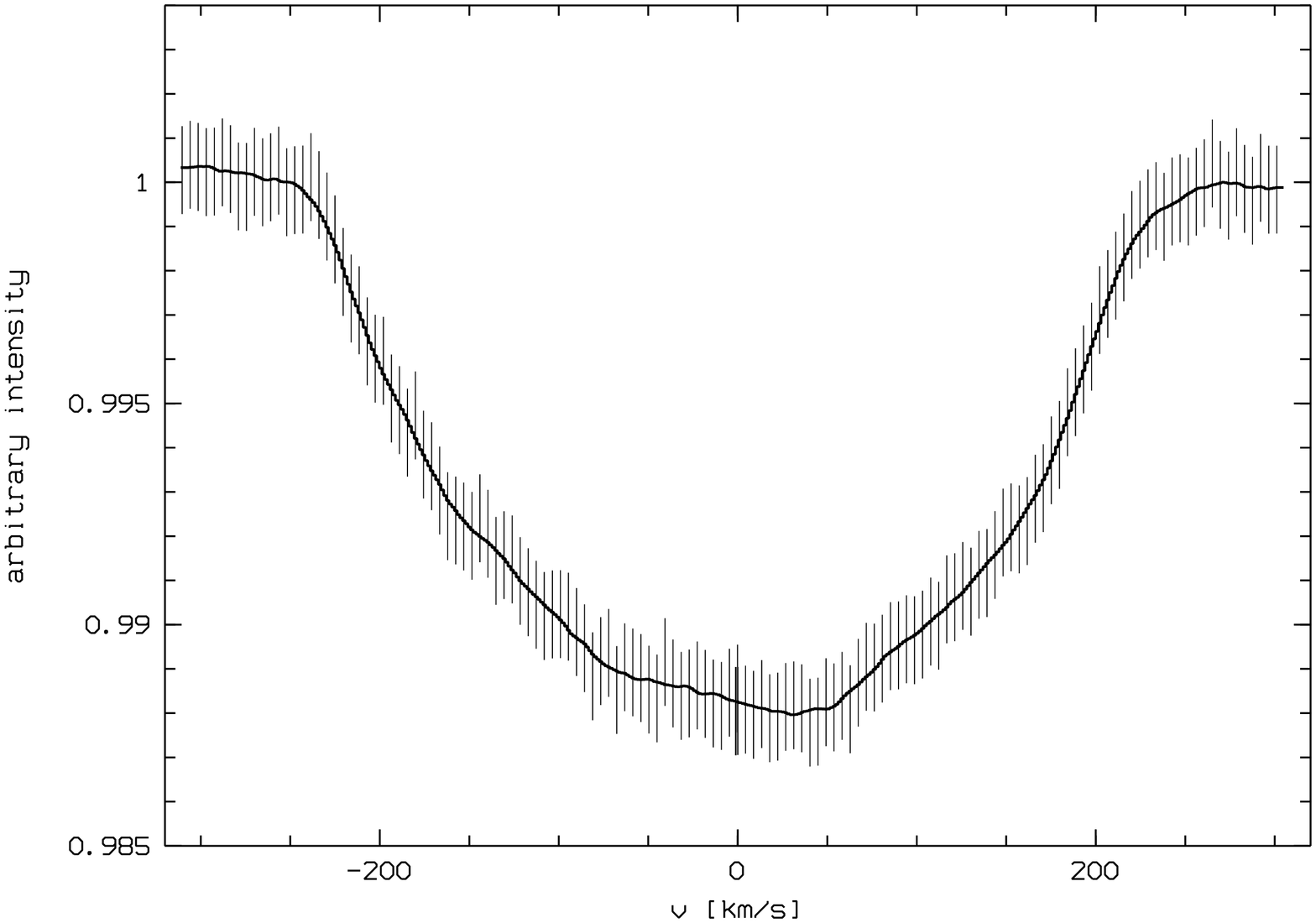}}
    \resizebox{.5\hsize}{!}{\includegraphics[angle=-90]{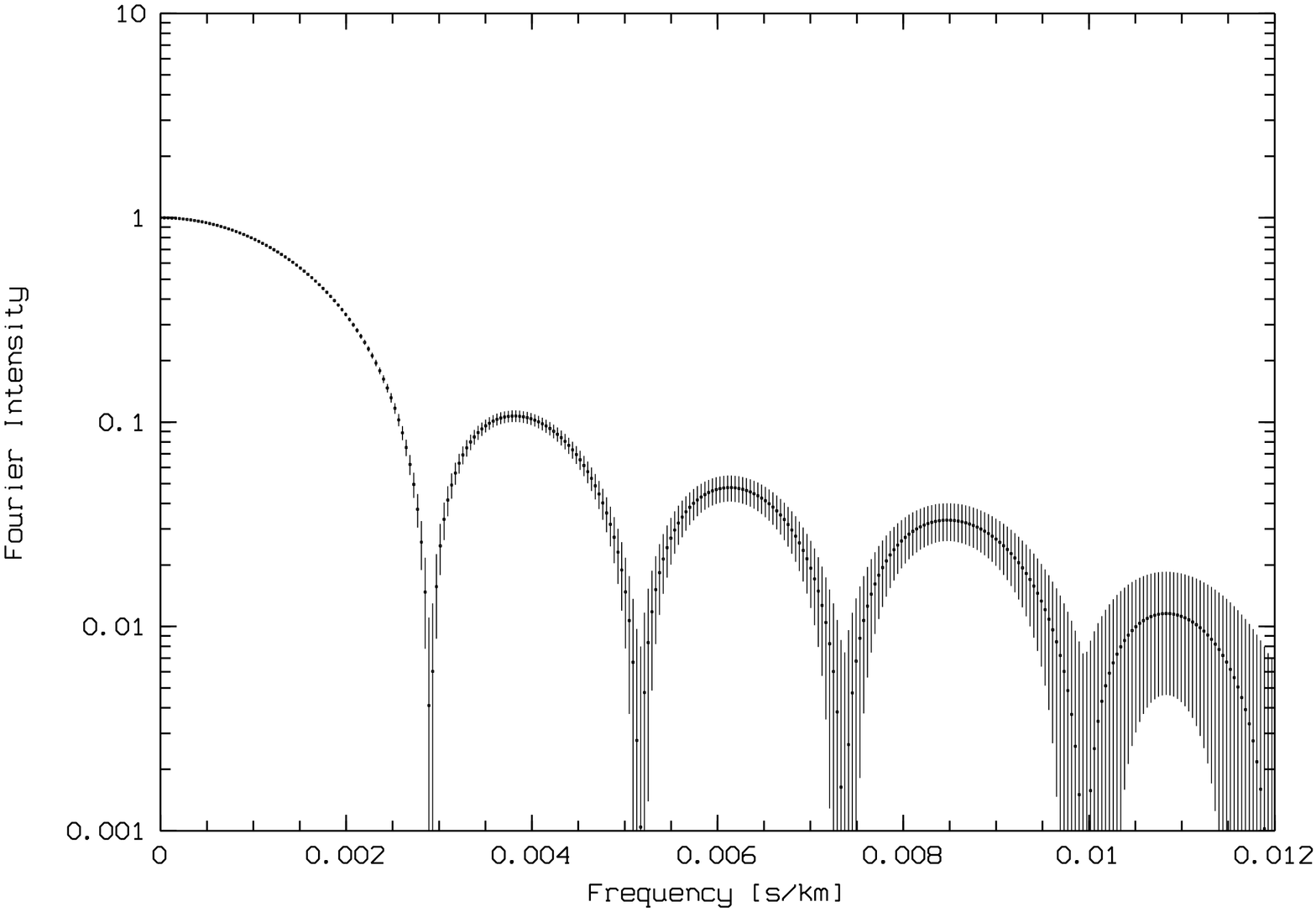}}}
    \caption{\label{plot:PSF} Left panel: global broadening function
      of Altair reconstructed from about 650 lines. Error bars
      according to a signal-to-noise ratio of 1000 are overplotted. 
      Right panel: Fourier transform of the reconstructed broadening
      function (left panel) with its error bars. Note the logarithmic
      scale on the y-axis.}
\end{figure*}

A small region of the result is shown in Fig.\,\ref{plot:Altair_fit},
fit quality is representative for the whole wavelength region. The
derived broadening function is shown in the inset. Note that no
additional changes were made to the continuum and that the broadening
function is nicely symmetric. A small linear slope in the overall
appearance might be visible. While this slope is likely due do tiny
continuum mismatch, however, it does not significantly affect the line
shape. The pixel size used is only a fraction of the instrument's
resolution and the small jitter in the broadening function is
numerical.  Taking the jitter as signal-to-noise ratio (S/N) yields
values of several thousands. We calculated the variance of each pixel
in the chi-square fit with regard to the two nearest neighbors of
each pixel. This also yields a S/N of several thousands.
Unfortunately, uncertainties in the broadening function are dominated
by our knowledge of the continuum and by systematically incorrect line
strengths (in detail, line strengths of Altair definitely differ to
those of HD~27\,819).  For the calculation in what follows we estimate
a rather conservative S/N of 1000 as the ``$1\sigma$''-level.
Consequently we use a S/N of 500 for our calculations of $2\sigma$
borders. In the left panel of Fig.\,\ref{plot:PSF} the global
broadening function of Altair is plotted with error bars according to
a S/N of 1000.

The global broadening function shows small asymmetries. From our
analysis we can not say whether these asymmetries are real or due to
incorrect line strengths in the template. However, within the error
bars the broadening function is symmetric, i.e. the broadening
function mirrored at the center falls withing the error bars.

In the right panel of Fig.\,\ref{plot:PSF} the Fourier transform of
the derived broadening function is shown with the respective $1\sigma$
error bars. The signal can clearly be followed up to the third
sidelobe before it reaches noise level. In the calculation of
$v\,\sin{i}$ and $i$ only the positions of the first two zeros are
needed; apparently both are well defined.

\section{Spectroscopic rotation and inclination}

Studying the characteristics of rotational broadening, zeros of the
broadening profile's Fourier transform are convenient observables
\citep[cp.][]{Gray92}.  Since convolutions in Fourier domain become
multiplications, especially zeros inferred from any broadening
mechanism remain unchanged by other effects like turbulence etc.; the
intensity is changed -- the position of the zeros is not. In case of
a star rotating as fast as Altair, no effects are expected inferring
signatures at frequencies as low as those inferred by rotation. Thus,
measuring Fourier transform zeros is a very robust method to access
rotational broadening -- and the zeros can easily and precisely be
determined.  The first zero is frequently used to determine stellar
projected rotational velocities $v\,\sin{i}$, as shown in
\citet{Dravins90} for the case of linear limb darkening.

Effects of rotation and inclination on absorption line profiles have
been studied by \cite{Reiners03a}. He shows that the ratio $q_2/q_1$
of the first two zeros $q_1$ and $q_2$ is sensitive for very fast
rotation. He concludes that the ratio $q_2/q_1$ does not depend on the
value of $v\,\sin{i}$, but on the equatorial velocity $v$ regardless
of the inclination angle $i$ (and on the rotation law, see
Sect.\,\ref{DiffRot}). Thus measuring $v\,\sin{i}$ from $q_1$ and $v$
from the ratio $q_2/q_1$ yields the value of the inclination angle
$i$!

By modeling absorption profiles of fast rotators, \citet{Reiners03a}
finds that the ratio $q_2/q_1$ can be expressed as
\begin{equation}
  \label{eq:Polynomial}
  q_{2}/q_{1} = 1.75 + av + bv^2.
\end{equation}
Values of $a$ and $b$ are given for different stellar models in
Table\,1 of that paper. The mechanism determining the values of $a$
and $b$ is gravity darkening described by the parameter $\beta$ in
\begin{equation}
\label{Eq:GravDarkLaw}
  T_{\rm eff} \propto g^{\beta},
\end{equation}
with $g$ the stellar surface gravity.  For Altair we get a value of
$\beta \sim 0.09$ from the tables given in \citet{Claret98}, i.e. the
results calculated by \citet{Reiners03} for the case of an F0 star are
directly applicable to Altair. Thus, for the parameters in
Eq.\,\ref{eq:Polynomial} we find $a = .740\cdot10^{-4}$ and $b =
-.345\cdot10^{-6}$.

Our measurements of the first two zeros $q_1$ and $q_2$ of the
broadening profile's Fourier transform are given in
Table\,\ref{tab:Results}.  Assuming a linear limb darkening law with a
limb darkening parameter $\epsilon = 0.6$ we calculate Altair's
projected rotational velocity as $v\,\sin{i} = 227 \pm
2$\,km\,s$^{-1}$.  Note that the formal error due to the uncertainty
of $q_1$ is given. We estimate the systematic error in $v\,\sin{i}$ to
be about 5\% due to continuum and line strength uncertainties
\citep[cp.][]{Reiners03}. From Eq.\,\ref{eq:Polynomial} we can
calculate Altair's equatorial velocity $v$ from the ratio $q_2/q_1$.

\begin{figure*}
  \centering \mbox{
    \resizebox{.5\hsize}{!}{\includegraphics{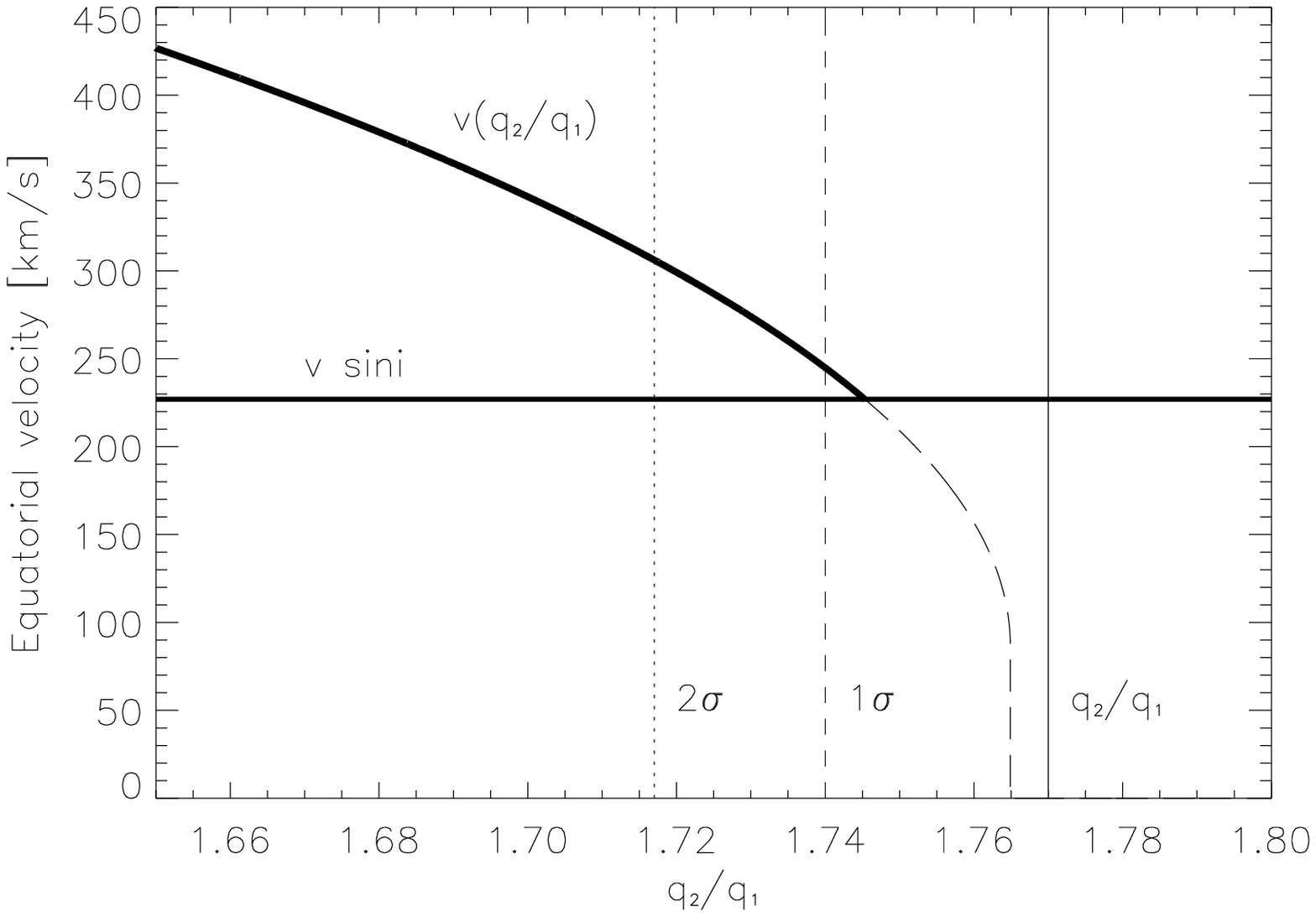}}
    \resizebox{.5\hsize}{!}{\includegraphics{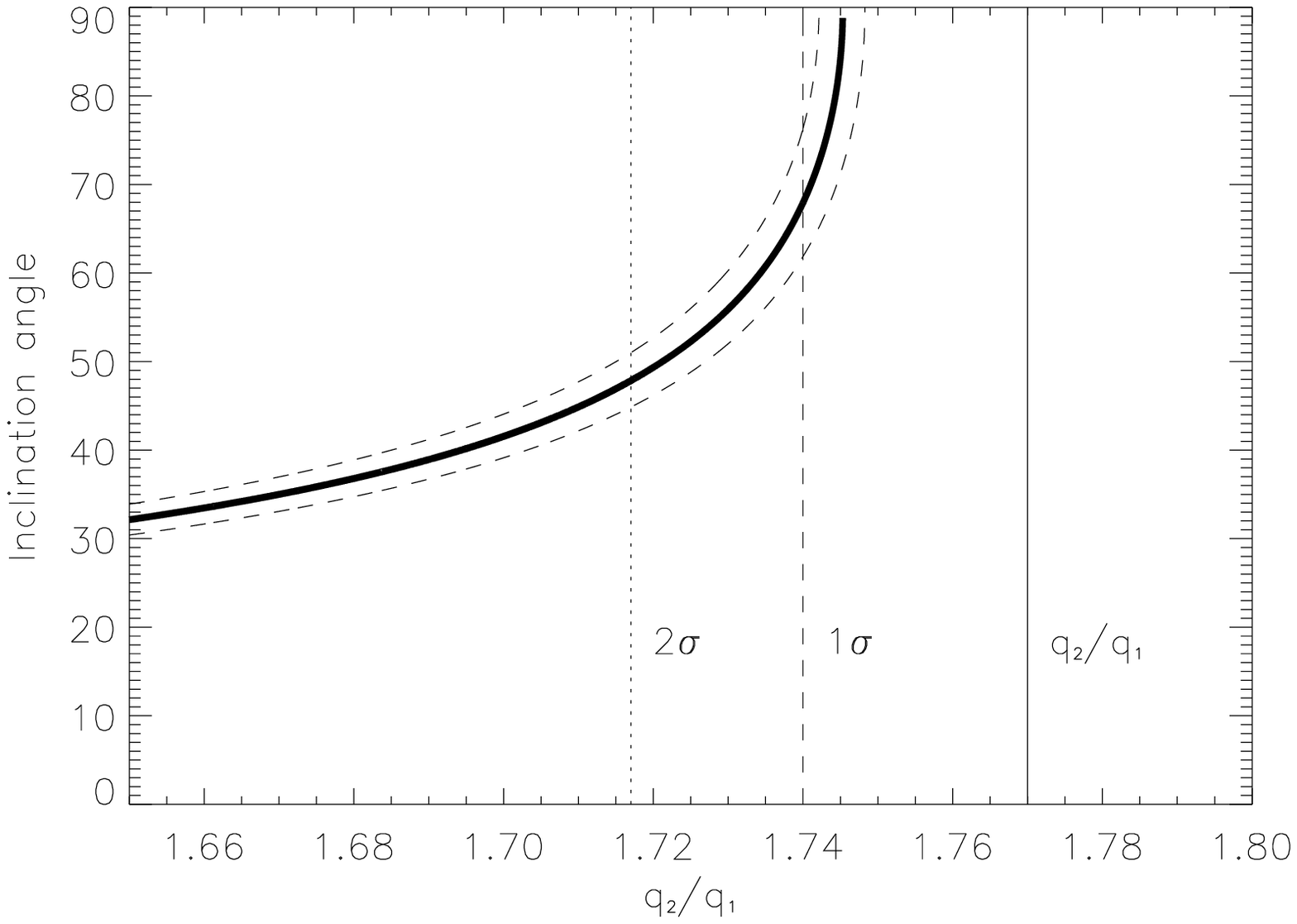}}}
  \caption{\label{plot:inclination} Left panel: Equatorial velocity 
    $v$ as a function of $q_2/q_1$ according to
    Eq.\,\ref{eq:Polynomial}.  Vertical lines indicate measured values
    of $q_2/q_1$ and its confidence levels, the horizontal lines shows
    the measured $v\,\sin{i}$. The allowed region where $v >
    v\,\sin{i}$ is plotted as a thick line. Right panel: Inclination
    angle $i$ (in degree) of Altair as function of $q_2/q_1$ (solid
    thick line).  Measurements of $q_2/q_1$ are plotted as in left
    panel. Dashed lines indicate the region allowed taking into
    account the systematic errors of $v\,\sin{i}$.}
\end{figure*}

In the left panel of Fig.\,\ref{plot:inclination} $v$ is plotted as a
function of $q_2/q_1$ with our measurement of $v\,\sin{i}$ overplotted
as a horizontal line, and the measurement of $q_2/q_1$ overplotted
with its $1\sigma$ and $2\sigma$ error levels as vertical lines. The
plot shows, that $q_2/q_1$ is generally expected being smaller than
1.765, i.e. the measured value of $q_2/q_1 = 1.77$ is not expected for
any possible inclination angle. However, within the $1\sigma$ region
we find possible values of $v(q_2/q_1)$ which are also consistent with
the measured value of $v\,\sin{i}$.

Calculating Altair's inclination angle $i$ from the measured
$v\,\sin{i}$ and $v(q_2/q_1)$ we find $i(q_2/q_1)$. We plot
$i(q_2/q_1)$ in the right panel of Fig.\,\ref{plot:inclination}.
$1\sigma$ and $2\sigma$ confidence levels in $q_2/q_1$ and systematic
errors of 5\% in $v\,\sin{i}$ are indicated by dashed lines. We find
that our measurements of $v\,\sin{i}$ and $q_2/q_1$ are consistent
within the $1\sigma$ level for values of $i > 68^{\circ}$.

\subsection{Intrinsic line variations}
\label{Temperature}
Variations of line profiles and line strengths have not been taken
into account in our procedure of line profile reconstruction. For the
case of Vega (A0V), \citet{Gulliver94} showed that for the two lines
they used line intrinsic effects dominate the line profile; they
observe a flattened profile instead of a cuspy shape. The difference
to our method is twofold: (i) gravity darkening as expressed in terms
of the parameter $\beta$ (Eq.\,\ref{Eq:GravDarkLaw}) is only 0.09 for
our case of Altair while it is 0.25 for the case of Vega. Temperature
variations with gravity are thus much weaker in the case of Altair;
(ii) our reconstruction procedure uses about 650 lines to determine
the shape of the line profile; we get a weighted mean of 650 line
profiles. Although we do not know specific temperature dependencies
for all lines, we do not expect them to be systematic.  Studying
temperature and gravity dependence of line strengths and shapes for
different spectral types would significantly reduce the systematic
errors but is beyond the scope of this paper. For this study we
estimate the systematic errors as given below and leave the detailed
study for a forthcoming publication.

\subsection{Differential rotation?}
\label{DiffRot}
It has been argued that Altair might undergo substantial surface
differential rotation since it was found to be an X-ray source
\citep{Ferrero95}. The parameter $q_2/q_1$ is also a sensitive
indicator for solar-like surface differential rotation with the
equator rotating at higher velocity than the polar regions do. In the
case of differential rotation, $q_2/q_1$ is expected to be
significantly smaller that 1.75 which is not the case for Altair. We
thus conclude that no signatures of solar-like differential rotation
could be found in the line profiles of Altair, i.e. solar-like
differential is not stronger than 10\% \citep{Reiners02}. However,
Altair is at least two orders of magnitude fainter in X-rays than
active F-stars are. Even if a magnetic dynamo was the source for this
faint X-ray flux, differential rotation may be below the detection
limit.

\section{Conclusions}

\begin{table}
  \caption{ \label{tab:Results} Measured values of Fourier transform zeros and calculated parameters $v\,\sin{i}$, $v$ and $i$. For $v\,\sin{i}$ formal and estimated systematic errors are given. For $1\sigma$ and $2\sigma$ confidence levels in $v$ and $i$, signal-to-noise ratios of the broadening function of 1000 and 500 are adopted, respectively.}
  \begin{tabular}{cc}
    \hline
    \hline
    \noalign{\smallskip}
    Parameter & Value\\ 
    \noalign{\smallskip}
    \hline
    \noalign{\smallskip}
    $q_1$ & $(2.91 \pm 0.03)\cdot10^{-3}$\,s\,km$^{-1}$\\
    $q_2$ & $(5.16 \pm 0.08)\cdot10^{-3}$\,s\,km$^{-1}$\\
    $\frac{q_2}{q_1}$ & $1.77 \pm 0.03$\\  
    \noalign{\smallskip}
    \hline
    \noalign{\smallskip}
    $v\,\sin{i}$   & $(227 \pm 2 \pm 11)$\,km\,s$^{-1}$\\
    \noalign{\smallskip}
    $v$ & $< 245$\,km\,s$^{-1} \quad (1\sigma)$\\
    & $< 305$\,km\,s$^{-1} \quad (2\sigma)$\\
    \noalign{\smallskip}
    $i$ & $>68^{\circ} \quad (1\sigma)$\\
    & $>45^{\circ}     \quad (2\sigma)$\\
    \noalign{\smallskip}
    \hline
  \end{tabular}
\end{table}

We have spectroscopically determined constraints on Altair's
inclination angle and differential rotation from the global rotational
broadening profile derived from about 650 spectral absorption lines.
The measured observables and our results are summarized in
Table\,\ref{tab:Results} where values of equatorial velocity
$v$ and inclination angle $i$ are given with their
respective $1\sigma$ and $2\sigma$ levels.

No signatures of surface differential rotation were detected, i.e.
solar-like differential rotation can be ruled out to be stronger than
10\% \citep{Reiners02}.

From the measured zeros of the Fourier transformed broadening profile
we conclude that Altair's equatorial velocity is not faster than
245\,km\,s$^{-1}$ (305\,km\,s$^{-1}$) or 53\,\% (72\,\%) of breakup
velocity $v_{\textrm{crit}} = 426 \pm 12$\,km\,s$^{-1}$
\citep{vanBelle}.  Thus its rotation axis is seen under an inclination
angle larger than $68^{\circ}$ on a $1\sigma$-level ($45^{\circ}$ on a
$2\sigma$-level). These results put additional constraints on the
possible rotation rate and inclination angle of Altair as determined
from interferometric data \citep{vanBelle}.

%

\end{document}